\documentclass{appolb}
\usepackage{epsfig}
\usepackage{amsfonts, amssymb, amsmath, graphicx, comment, bm, slashed, caption, subcaption, dcolumn,color}
\newcommand{\beq}{\begin{equation}}
\newcommand{\eeq}{\end{equation}}
\newcommand{\calH}{ {\cal H} }

\newcommand{\rmi}{ {\rm i} }

\begin{document}
\title{QCD in stars
\thanks{Presented at CPOD2016 in Wroclaw, Poland.}
}
\author{Toru Kojo
\address{Key Laboratory of Quark and Lepton Physics (MOE) and Institute of Particle Physics, Central China Normal University, Wuhan 430079, China}
}
\maketitle
\begin{abstract}
We discuss cold dense QCD by examining constraints from neutron stars, nuclear experiments, and QCD calculations at low and high baryon density. The two solar mass constraint and suggestive small radii ($\sim 10-13$ km) of neutron stars constrain the strength of hadron-quark matter phase transitions. Assuming the adiabatic continuity from hadronic to quark matter, we use a schematic quark model for hadron physics and examine the size of medium coupling constants. We find that to baryon density $n_B \sim 10n_0$ ($n_0$: nuclear saturation density), the model coupling constants should be as large as in the vacuum, indicating that gluons remain non-perturbative even after the quark matter formation.
\end{abstract}
\PACS{PACS numbers come here}

\section{Introduction}

The relativistic heavy ion collisions at RHIC and LHC, together with the lattice Monte-Carlo simulations, allow us quantitative studies of the phase diagram at high temperature and low baryon density. There the QCD phase transition is known to be a crossover in which a hadronic matter continuously transforms into a quark-gluon plasma. Then it also raises a question on the nature of hadron-quark phase transitions at low temperature and high baryon density. The questions addressed in this talk are: (i) what is the nature of hadron-quark matter phase transitions? (ii) how does the gluon dynamics change as density increases? (iii) what kind of many-body correlations emerge at high baryon density? We try to get some hints to answer these questions by studying equations of state which are nowadays strongly constrained by neutron star observations \cite{2m_1,Ozel2010,Steiner:2012xt}, nuclear physics \cite{Danielewicz:2002pu,Steiner:2004fi}, and QCD calculations at low \cite{Gandolfi:2015jma} and high \cite{Kurkela:2009gj} density limits.

\begin{figure}[h]
\vspace{-0.2cm}
\begin{minipage}{0.57\hsize}
\vspace{0.1cm}
\hspace{-0.1cm}
\includegraphics[width = 0.77\textwidth]{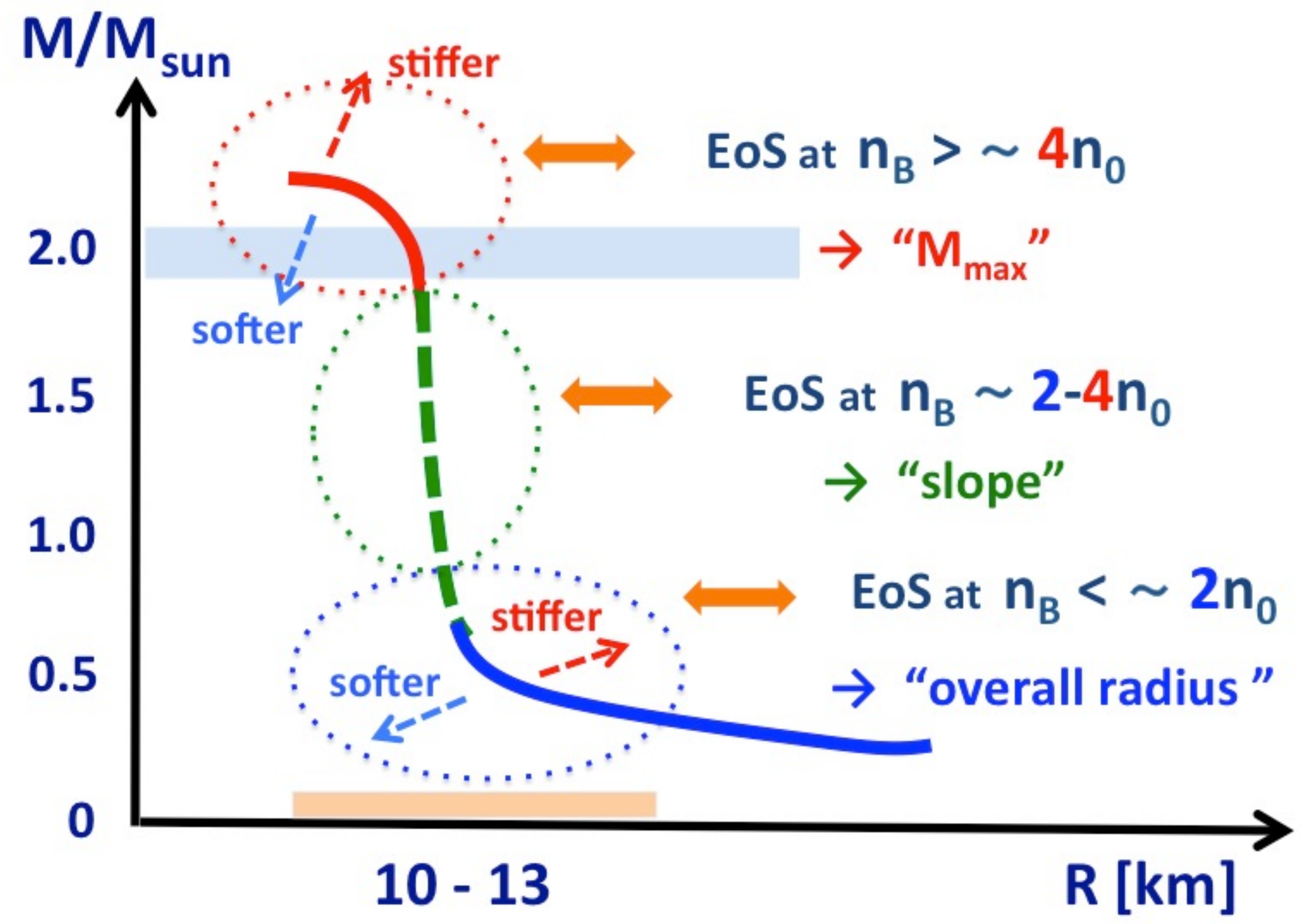}
\end{minipage}
\begin{minipage}{0.57\hsize}
\vspace{0.1cm}
\hspace{-0.4cm}
\includegraphics[width = 0.77\textwidth]{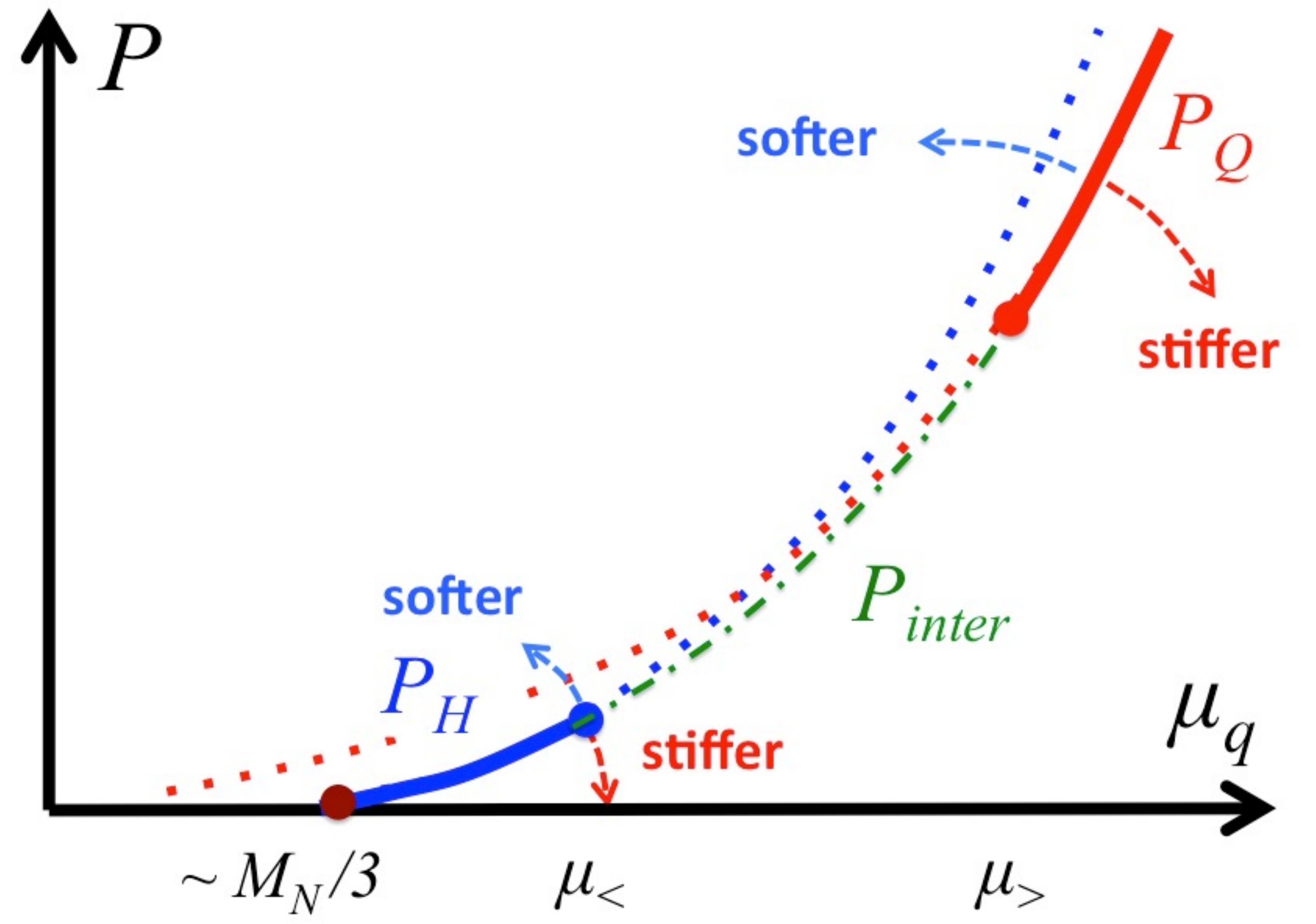}
\end{minipage}
\vspace{-0.cm}
\caption{
\footnotesize{
(left) The correlation between the shape of the $M$-$R$ curve and pressures at several fiducial densities. (right) $P(\mu_q)$ curves. The 3-window modeling assumes that only the bold lines is trustable (the dotted lines are their extrapolations). The green curve is the interpolated pressure.
}
}
\vspace{-0.3cm}
\label{MR}
\end{figure}

\section{Constraints from theories and observations}

There are theoretical guides from QCD calculations. At low density, realistic nuclear potentials, combined with the sophisticated many-body calculations \cite{Gandolfi:2015jma}, are able to explain the properties of light nuclei and nuclear matter in good accuracy. At $n_B > 2n_0$, however, there arise a number of conceptual questions such as the convergence of many-body forces, hyperon softening problem, and the possibility of the structural changes in hadrons. The validity of hadronic calculations should be understood from microscopic treatments of QCD dynamics.

At high density, perturbative QCD (pQCD) calculations have been carried out to 3-loop order \cite{Kurkela:2009gj}. They suggest that the expansion works at quark chemical potential $\mu_q$ larger than $\sim 1\, {\rm GeV}$ or at baryon density $n_B \sim 100n_0$, while at lower density the results show the large renormalization scale dependence, indicating that soft gluons are important and the matter is strongly correlated. 

The domain $2n_0 \lesssim n_B \lesssim 100 n_0$ is hard to explore directly from QCD calculations, but neutron star observations provide us with considerable information. In principle, if the $M$-$R$ relation for neutron stars are established, one can use it to directly reconstruct equations of state \cite{Lindblom1992}. Such direct conversion is not possible at present, but the current data can already impose significant constraints on possible equations of state. 

Practically it is very useful to note that the shapes of mass-radius ($M-R$) curves are roughly determined by the pressure at several fiducial densities \cite{Lattimer:2006xb} (Fig.\ref{MR}, left). The overall radii of typical neutron stars are determined by the pressure at $n_B \sim 2n_0$, and softer (stiffer) equations of state give smaller (larger) radii. The recent trend obtained from the (suggestive) estimates of neutron star radii with $R= 10-13\,{\rm Km}$  for $M\sim 1.4 M_{\odot}$ ($M_\odot$: solar mass) \cite{Steiner:2012xt,Ozel2015} indicates soft equations of state at $n_B \sim 2n_0$, and this is consistent with heavy ion data \cite{Danielewicz:2002pu}, nuclear symmetry energy \cite{Steiner:2004fi}, and recent advance quantum Monte-Carlo calculations \cite{Gandolfi:2015jma}. Further constraints will be obtained if the gravitational waves from neutron star mergers are discovered \cite{Baiotti:2016qnr}. With equations of state at $n_B > 2n_0$, the $M-R$ curves start to rise almost vertically and the slopes are determined by pressure for $2n_0 \lesssim n_B \lesssim 5n_0$. Then the curves reach their maximal masses whose values strongly correlate with equations of state at $n_B \gtrsim 5n_0$. The existence of $2M_\odot$ stars play significant roles here: it requires that equations of state at $n_B \gtrsim 5n_0$ must be very stiff, and many model equations of state were already rejected by this condition \cite{2m_1}.

The softness at low density and stiffness at high density together give significant constraints on possible equations of state for $2n_0 \lesssim n_B \lesssim 5n_0$. In general, such soft-to-stiff behaviors require the increase of pressure for finite energy interval, but such increase must be sufficiently slow in order to keep the causality constraint, $c_s^2 = \partial P/\partial \varepsilon \le c^2$, where $c_s$ is the (adiabatic) speed of sound and $c$ is the light velocity. The restriction becomes even severer if we have the strong first order phase transition in the intermediate region; because of the softening associated with first order phase transitions, after the transitions we must introduce rapid stiffening within small energy density interval to get the connection with stiff high density equations of state. For these reasons, we are inclined to think that the intermediate region does not have strong first phase transitions, and low and high density matter are connected by smooth crossover or weak first order transitions\footnote{If the equations of state at $n_B\sim 2n_0$ turns out to be very stiff, then we can still allow strong first order phase transitions \cite{Benic:2014jia}. }.

These theoretical and observational constraints bring us to the 3-window description of dense matter \cite{Masuda:2012kf,Kojo:2014rca}: (i) At $n_B \lesssim 2n_0$, the matter is dilute and nucleons exchange only few mesons so that baryons and mesons are well-defined objects. (ii) At $n_B \gtrsim 2n_0$, baryons start to exchange many mesons (or quarks) and many-body forces become increasingly important. With many quark exchanges the identity of a hadron becomes ambiguous. (iii) At $n_B \gtrsim 5n_0$, baryons overlap and quarks begin to develop the quark Fermi sea, forming a quark matter. Nevertheless the matter is supposed to be strongly coupled according to the pQCD calculations.

We construct equations of state based on this picture (Fig.\ref{MR}, right). At $n_B < 2n_0$, we use Akmal-Pandharipande-Ravenhall (APR) equation of state\footnote{For the crust part we use the SLy equations of state \cite{Douchin:2001sv} up to $n_B = 0.5n_0$. } as a representative of nuclear equations of state \cite{Akmal:1998cf}. For $n_B > 5n_0$, we use a schematic quark model for $n_B > 5n_0$ leaving the in-medium coupling constants as free parameters. The $2M_\odot$ constraint limits the range of the effective couplings and from which we delineate the properties of dense matter \cite{Kojo:2014rca}. The most difficult is the modeling of matter at $2n_0 < n_B < 5n_0$, where neither purely hadronic nor quark matter descriptions are reliable. Thus we construct the equations of state by interpolating the APR and quark model pressure and applying thermodynamic and causality constraints. This patchwork study apparently has a lot of freedoms, but actually the recent constraints are so tight that one can derive useful insights into the QCD phase diagram.

\section{Delineating the properties of matter at high density}

Based on the crossover picture, we assume the adiabatic continuity in which the structure of effective interactions are smoothly connected from low to high density. We use the Nambu-Jona-Lasinio (NJL) model \cite{Hatsuda:1994pi} with effective interactions inspired from hadron and nuclear physics,
\begin{align}
\calH  
 =&~ \calH_{ {\rm NJL} }
+ \calH_{ {\rm conf} }^{ {\rm 3q\rightarrow B} } \nonumber \\
& - \frac{H}{2} \!\sum_{A,A^\prime = 2,5,7} \!
 \left(\bar{q} \rmi \gamma_5 \tau_A \lambda_{A^\prime} C \bar{q}^T \right) \left(q^T C \rmi \gamma_5 \tau_A \lambda_{A^\prime} q \right) + \frac{G_V}{2} (\overline{q} \gamma^\mu q)^2
 \, ,\nonumber 
\end{align}
where $\tau_A$  and $\lambda_A$ are Gell-Mann matrices for flavors and colors respectively. The Hamiltonian $\calH_{ {\rm NJL} }$ is the standard NJL model which is responsible for the descriptions of chiral symmetry breaking/restoration and the changes in quark bases. The Hamiltonian $\calH_{\rm conf}^{ {\rm 3q\rightarrow B} }$ is responsible for the confining forces that trap three quarks into a baryon. This term should be crucial at low density but we will drop it off by restricting the use of our model to the quark matter domain. The third term expresses the color magnetic interaction which is responsible for the $N-\Delta$ splitting and the color super conductivity at high density. The last term is the repulsive density interaction which is inspired from the repulsive $\omega$-meson exchange in nuclear physics.

The {\it in-medium coupling constants}, $H$ and $G_V$, at $n_B >5n_0$ are treated as free parameters while we fix the scalar coupling constant $G_s$ in $\calH_{ {\rm NJL} }$ to the vacuum value\footnote{In principle we should also vary $G_s$ but we found that $G_s$ at $n_B>5n_0$ should be as large as its vacuum value to avoid strong first order phase transitions. }, $G_s=G_s^{ {\rm vac} }$. We note that the effective interactions summarize the microscopic processes of quarks and gluons in a concise way, and their importance is reflected in the values of $(G_s, G_V, H)$. If the matter is weakly coupled and gluons become perturbative, the equations of state should be like an ideal gas so that $(G_s, G_V, H)$ should be much smaller than the vacuum values. Instead, our model analyses suggest that the effective couplings at $n_B \gtrsim 5n_0$ should be comparable to the vacuum value to pass the $2M_\odot$ constraint. In short, we use $G_V \sim G_s^{ {\rm vac}}$ to stiffen equations of state, which in turn requires $H\sim G_s^{ {\rm vac}}$ for the smooth connection to the APR equation of state at $n_B=2n_0$. In Ref. \cite{Fukushima:2015bda}, more elaborated treatments of the running $G_V$ were performed.

This picture is actually consistent with our implicit assumption on the QCD vacuum energy. For instance suppose that the matter at $n_B \sim 5n_0$ becomes weakly coupled and perturbative. Then the resulting pressure must consist of perturbative contributions {\it plus a constant term}, $B$ (bag constant), that reflects the difference between the perturbative and non-perturbative vacua. The size of such bag constant is naively given by the QCD scale $\sim (200\,{\rm MeV})^4$, comparable to the energy density at the core of neutron stars. If such large bag constant would appear, it would reduce the pressure by $B$ while increase the energy density by $B$, causing the significant softening in the equations of state. Thus we expect that the gluons remain non-perturbative and do not produce significant bag constant at $n_B \sim 5n_0$.

In summary, we delineate properties of QCD matter by studying the equations of state supposed from theoretical and empirical constraints. It seems to us that the crossover phase transition together with adiabatic continuity from hadronic to quark sectors offer a consistent description of QCD matter. Gluons seem non-perturbative even after the quark matter formation. Such regime has been studied in the context of quarkyonic matter in which one takes the large $N_{{\rm c}}$ limit to keep gluons in medium as non-perturbative as in the QCD vacuum \cite{McLerran:2007qj}. At finite $N_{{\rm c}}$, the description of gluons in neutron star densities is a highly nonlinear problem; it strongly depends on how strongly quarks near the Fermi surface fluctuate, but the nature of such fluctuations is very sensitive to the condensed phases in consideration \cite{Rischke:2000cn}. The effects of the condensates appear in the excitation modes \cite{Bedaque:2001je} and the resulting thermal equations of state. The future detection of gravitational waves from neutron star mergers will provide us with such information, with which we should be able to discriminate various phase structures and test theoretical scenarios.

I thank P. Powell, Y. Song, G. Baym, and K. Fukushima for the collaborations. I would like to express my gratitude to the workshop organizers for inviting me to this enjoyable workshop.


\end{document}